\documentclass[12pt]{iopart}

\usepackage{graphicx}

\usepackage{color}

\begin{document}

\title[]{Scanning electron microscopy of cold gases}

%\author{IOP Publishing$^1$\footnote{Corresponding 
%author: esubs@iop.org} and (modified by) Mitsuaki 
%Funakoshi$^2$\footnote{Corresponding author: fdr@acs.i.kyoto-u.ac.jp}}

\author{Bodhaditya Santra and Herwig Ott}

\address{Department of Physics and Research Center OPTIMAS,Technische Universit{\"a}t Kaiserslautern, 67663 Kaiserslautern, Germany}

%\ead{graham.douglas@iop.org}

\begin{abstract}
Ultracold quantum gases offer unique possibilities to study interacting many-body quantum systems. Probing and manipulating such systems with ever increasing degree of control requires novel experimental techniques. Scanning electron microscopy is a high resolution technique which can be used for {\it in situ} imaging, single site addressing in optical lattices and precision density engineering. Here, we review recent advances and achievements obtained with this technique and discuss future perspectives.
\end{abstract}

\maketitle

\tableofcontents

\section{Introduction}

Scientific breakthroughs has always been associated with technological progress. The emergence of more sophisticated measurement techniques enables the discovery of new and fascinating phenomena. In particular the exploration of microscopic structures greatly benefits from the ever growing capabilities in imaging objects with high spatial resolution. At an extremely small length scale, the underlying quantum-mechanical nature eventually becomes observable. The realization of Bose-Einstein condensation in dilute atomic gases \cite{Anderson_1995,Bradley_1995,Davis_1995,Cornell_2002,Ketterle_2002} and early experiments on it \cite{Interference_Ketterle, Vortex_Ketterle} have demonstrated that at sufficiently low temperatures the quantum effects are `magnified' such that they become observable in real space \cite{Ketterle_1999,Ketterle_2008}. Since then, an explosion of theoretical and experimental work has set in, including strongly interacting lattice gases \cite{Bloch_2005}, interacting Fermi gases \cite{Chin_2004,Partridge_2006,Nascimbene_2010}, spinor systems \cite{Stamper-Kurn_2013}, artificial gauge fields \cite{Lin_2009, Lin_2011}, quantum optical applications \cite{Ginsberg_2007}, and quantum information processors \cite{Brennen_1999,Raussendorf_2001}.

Obviously, the desire for an in-depth understanding of the numerous phenomena, such as superfluidity, quantum correlations, or strongly interacting many-body quantum phases drives the quest for novel preparation and imaging techniques. Inspired by the principles of scanning electron microscopy, a technique which is unrivaled in its spatial resolution compared to optical methods, we have developed an apparatus for probing ultracold atomic gases by means of electron-atom scattering~\cite{Gericke_2006,Wurtz_2008,Gericke_2008}. The potential of the technique was demonstrated by the first image of an atomic Bose-Einstein condensate with sub-micrometer spatial resolution~\cite{Gericke_2008}. Scanning electron microscopy of ultracold quantum gases is not only suited for high resolution imaging, but also for the preparation of engineered density distributions in optical lattice sites ~\cite{Wurtz_PRL_2009} and {\it in vivo} studies of many-body quantum dynamics \cite{Guarrera_correlations_2011,Barontini_2013}.

In this review article, we describe how the principles of scanning electron microscopy (SEM) can be applied to the manipulation and detection of ultracold atoms and what phenomena can be studied with this technique. The structure of the article is as follows: In the second section, we give a general motivation for high resolution probing of ultracold quantum gases and provide references to the literature for more details. Section 3 is of mainly technical nature and introduces the principles of the imaging procedure. The relevant electron-atom scattering mechanisms are discussed and the working principle of the microscope is explained. We give special emphasis to the technical compatibility of SEM when combined with a cold atom experiment. High resolution images of ultracold quantum gases and the demonstration of single site addressability close this section. Section 4 presents a selection of three experiments which highlight different aspects how quantum gases can be studied with SEM: the detection of pair correlations, dissipation induced quantum Zeno dynamics and electron microscopy of a Rydberg excited Bose-Einstein condensate. In the last section we discuss the perspectives and possible extensions of SEM technique to fermionic quantum gases and to the study of open macroscopic systems.

\section{High precision probing and manipulation of ultracold quantum gases}

The realization of Bose-Einstein condensation and Fermi degeneracy in cold gases of atoms and molecules has opened a new era in experimental quantum physics~\cite{Bloch_2008}. Experimental and theoretical progress were accompanied by the development of detection and manipulation techniques, such as absorption and phase contrast imaging or radio frequency spectroscopy \cite{Ketterle_1999}. High precision probing and manipulation techniques were sought from early on. They are not only indispensable for the study of many-body systems on the single atom level, but can also be used to investigate the dynamics, the interactions and the coupling to an environment in the quantum regime. In the following we give a few examples how experiments can benefit from high performance imaging and manipulation techniques.

One motivation comes from {\it quantum simulation} with cold atoms in optical lattices. A completely new class of experiments and physical investigations becomes possible when atoms in individual sites of an optical lattice can be probed locally. This gives access to many characteristic properties of many-body quantum phases~\cite{Bloch_2012} such as the site occupancy, the tunneling dynamics and the nearest neighbor correlations. Electron microscopy \cite{Wurtz_PRL_2009} and high resolution optical microscopy ~\cite{Nelson_2007,Bakr_2009,Sherson_2010} has been developed over the past years and has demonstrated the huge potential of this approach. Related to these developments there are proposals for {\it quantum information processing}, based on arrays of single atoms in optical lattices~\cite{Raussendorf_2001}.

{\it Transport of particles} plays a central role in solid state systems. For example, the conductivity of a material depends on the quantum-mechanical transport properties of the electrons when moving through the microscopic structure of the material. With the help of cold atom systems, fundamental aspects of such transport processes can be studied under controlled conditions \cite{Brantut_2012,Ronzheimer_2013}. Tunneling is a common ingredient in transport processes and a high resolution imaging technique is ideally suited to characterize the microscopic physics and dynamics at a tunneling junction. 

Understanding the {\it non-equilibrium dynamics} of closed interacting quantum systems lie at the heart of many-body physics~\cite{Polkovnikov_2011,Langen_2014,Eisert_2014}. How a closed system transits from an initial out-of-equilibrium state to a final equilibrium state is not yet clearly understood. The ability to design, control and measure tailored quantum systems on the microscopic level paves the way towards understanding local and global properties of non-equilibrium dynamics.

The evolution of an {\it open many-body quantum system} is governed by the coupling of the system with the environment. During the last decades several advances have been made in the study of environmentally induced phenomena like decoherence and decoherence-induced selection of preferred states~\cite{Diehl_2008,Raimond_2010}. Engineering the action of an environment as a local dissipative mechanism is an important tool for designing many-body quantum state. {\it In situ} addressability of quantum gases can therefore help to tailor the environmental influence~\cite{Barontini_2013}, eventually realizing complex decoherence-free subspaces and dissipative many-body attractor states.

The most common methods for high resolution probing and manipulation of ultracold quantum gases are based on optical techniques (for details, see~\cite{Ketterle_1999} and references therein). The spatial resolution is diffraction limited and thus depends on the numerical aperture of the optical system and the wavelength of the imaging light. For current experiments, the highest reported resolution is about 600\,nm and is realized by quantum gas microscopes~\cite{Bakr_2009,Sherson_2010}. This resolution is sufficient to probe a quantum gas with single atom/single lattice site resolution in the Hubbard model regime.

Many length scales in ultracold quantum gases can even be smaller than the typical spacing in an optical lattice, e.g. the healing length, the correlation length, the average interparticle distance, the size of a Rydberg wave function or the extent of the Wannier function. SEM of ultracold gases provides a technique with a spatial resolution down to 100\,nm which can work in bulk as well as in lattice systems. Moreover, it combines {\it in situ} and {\it in vivo} probing of quantum gases, mass sensitive detection of different atomic species and density engineering of ultracold gases.

\section{Scanning electron microscopy of ultracold quantum gases}

\begin{figure}[t!]
\begin{center}
\includegraphics[width=0.7\textwidth]{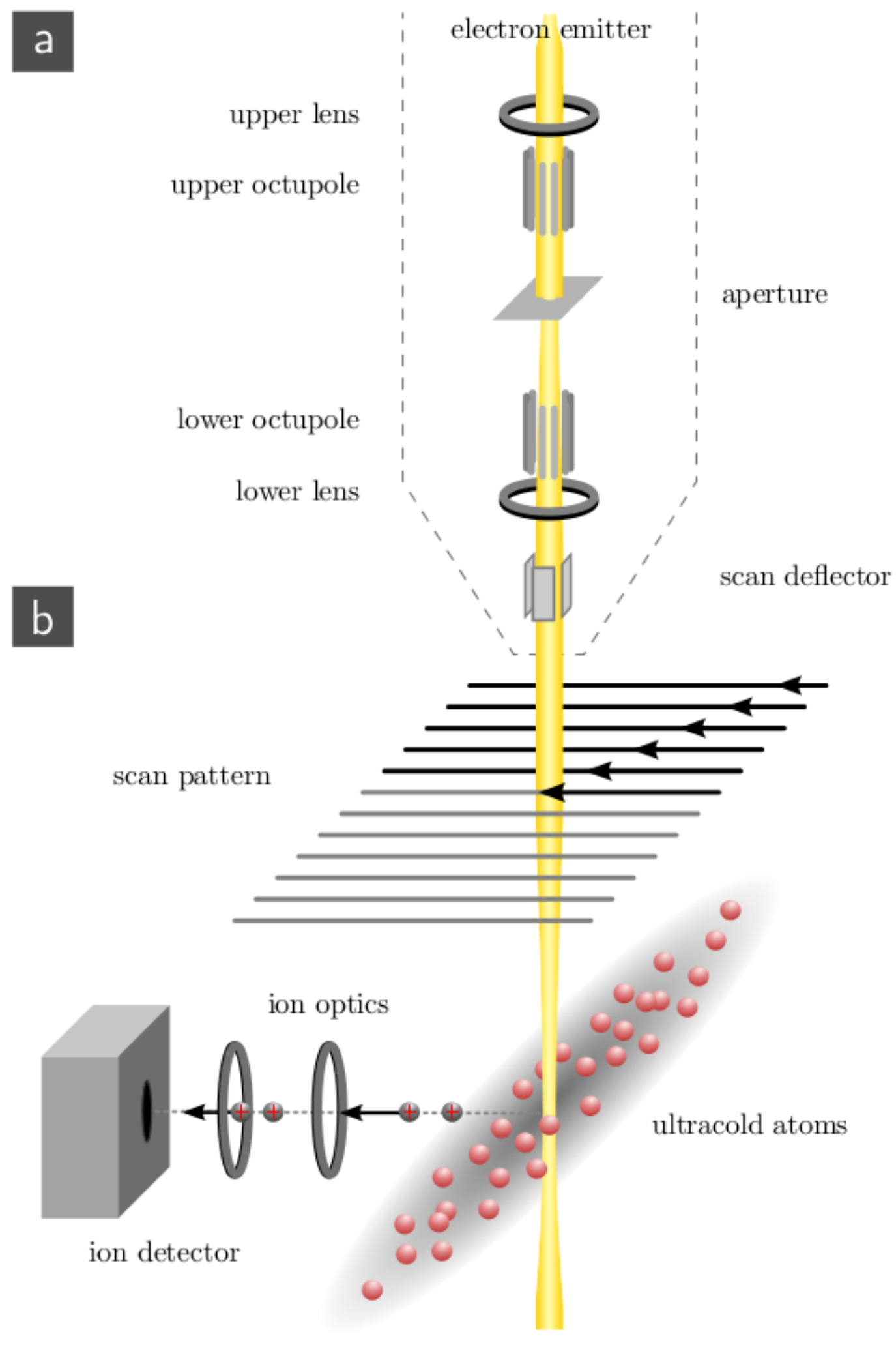}
\end{center}
\caption{Schematics of the working principle. (a) The electron column, represented by the dashed line, is based on two magnetic lenses and electrostatic deflection systems (see Fig.\,3 for more details). It provides a focused electron beam, which intersects the atomic sample, prepared in an optical dipole trap.  (b) The electron beam is scanned across the cloud. Electron impact ioization produces ions, which are guided with an ion optical system towards a channeltron detector. The ion signal together with the scan
pattern is used to compile the image. From Ref.~\cite{Wurtz_applPhysB2010}.} 
\label{wkprinciple} 
\end{figure}

This section contains a detailed description of the physical and technical basis of the imaging procedure. In the following paragraph, we summarize the most important aspects such that the reader can decide to skip the technical part and directly jump to the experimental results starting with section 3.3. 

The detection of ultracold atoms by scanning electron microscopy is based on electron impact ionization of the trapped atoms with subsequent ion detection (Fig.~\ref{wkprinciple}). The electron column provides a focused electron beam (EB) with 6 keV electron energy, a full-width-at-half-maximum (FWHM) diameter of 100-500 nm and a current of 10-500 nA. The smallest beam diameter we could achieve with the column is about 50\,nm. This size is determined by the electron optics, which are designed to magnify the virtual source size of the electron emitter (25\, - 30\,nm diameter) by a factor of 2.5. This magnification was chosen to allow for large beam currents. Different electron optics can produce much smaller beam diameters, however at the cost of beam current. The chosen configuration is a good compromise between resolution and imaging speed. The electron beam is scanned across an ultracold quantum gas of rubidium atoms, which is prepared in an optical dipole trap~\cite{Gericke_2007}. The ions produced are extracted with an electrostatic field and detected by a channeltron~\cite{Gericke_2008}. The small diameter of the electron beam ensures a high spatial resolution, whereas the ion detection provides single-atom sensitivity. The beam is scanned in a TV-like pattern and the temporal ion signal is assigned to the beam position in order to construct the image. A typical image sequence takes about 100\,ms, in which a few hundred atoms are detected. The overall detection efficiency is limited by the branching ratio between electron impact ionization and non-ionizing collisions and the detector efficiency. It amounts to 10 - 20 percent. As the cross-section for electron-atom scattering is eight orders of magnitude smaller than the absorption cross-section of a resonant photon, the atomic cloud is optically thin for the electron beam. For typical parameters, only one out of 500,000 incident electrons undergoes a collision. In-depth descriptions of the electron-atom interactions and the technical details are provided in the following, typical SEM images are provided in section 3.3.

\begin{figure}[t!]
\begin{center}
\includegraphics[width=0.7\textwidth]{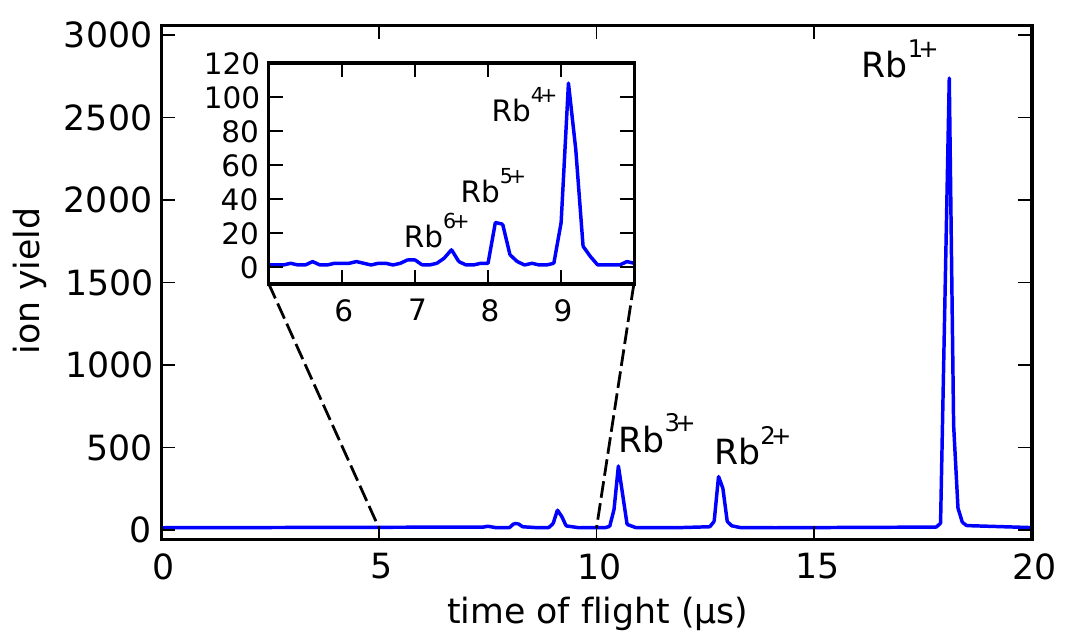}
\end{center}
\caption{Time-of-flight spectrum of $^{87}$Rb atoms ionized by electron impact ionization. 80$\%$ of the produced ions are singly charged. Higher charge states up to Rb$^{7+}$ are also observed. A peak width of 200 ns ensures that each detected event can be assigned to a spatial coordinate of the electron beam. From Ref.~\cite{Wurtz_applPhysB2010}.} 
\label{tof87Rb} 
\end{figure} 

\subsection{Interaction mechanisms}

In general, the interaction between an electron and a ground state atom can be elastic or inelastic. Electron impact ionization of atoms is a special case of inelastic scattering which we consider here as an independent process. In an elastic scattering process, the scattered atom carries some energy and momentum of the incident electron. The internal state of the target atom remains unchanged but leaves the trap due to the transferred momentum. In an inelastic scattering process, the target atom is excited to a higher state. The energy transfer to the atom is again accompanied by a momentum transfer. In both cases, the atoms escape from the trap and will not be detected as no ion is created. In electron impact ionization, the incident electron knocks out one or more bound electrons from the target atom leading to the formation of ions, which can be detected by an ion detector. It is this process that allows us to detect the atoms. Fig.~\ref{tof87Rb} shows a time-of-flight spectrum of the produced rubidium ions. About 80$\%$ of the ions are singly charged, whereas the remaining fraction is multiply charged. They originate from inner shell ionization events with subsequent autoionization cascades~\cite{Das_1985}.
 
The application of scanning electron microscopy to the detection of ultracold atoms has to take into account several constraints: the signal has to be large enough, the time resolution must be sufficiently high and the detection process should be efficient in order to avoid long measurement times. Multiple scattering events and secondary processes might additionally spoil the image and the detected ions have to be assigned to the correct position of the electron beam. The initial cold temperature of the atoms and the fact that the majority of the ionization processes occurs at small momentum transfer to the remaining ion ensure that the created ions have a negligible starting velocity. The time-of-flight to the detector is therefore identical for all singly charged ions (see Fig.\,\ref{tof87Rb}) and the detection time of the ion can be unambiguously assigned to the position of the electron beam~\cite{Wurtz_applPhysB2010}.

The signal strength is a crucial parameter. As any scattered atom is removed from the gas, the maximum signal recorded at the detector is given by the number of atoms in the gas multiplied with the ionization probability.

The total ionization cross-section for rubidium at 6 keV electron energy is $\sigma_{ion}=7\pm 0.7\times 10^{-17}$cm$^{2}$~\cite{Wurtz_2010} and represents 40$\%$ of all scattering events~\cite{Schappe_1995,Schappe_1996}. Elastic and inelastic electron-atom collisions constitute the remaining events and lead to atom loss with no detectable signal. The sum of all three contributions gives the total scattering cross section, which amounts to $\sigma_{tot} = 1.78\pm 0.14 \times 10^{-16}$ cm$^{2}$~\cite{Wurtz_2010}. A detailed description of the elastic and inelastic scattering processes and electron impact ionization can be found in reference~\cite{Coplan_1994}. An atom that is located in the center of the electron beam, has a lifetime against collisions with incoming electrons of 

\begin{equation}
\tau=\frac{e}{j_0 \sigma_{tot}},
\end{equation}
where $j_0$ is the current density in the beam center. For typical beam parameters we find $\tau$ = 5 $\mu s$, which gives an estimate for a reasonable dwell time per pixel. Most of the experiments are performed with a dwell time of 1 or 2 $\mu$s per pixel. This limits the overall number of detected atoms but allows for a fast scanning procedure. A scanning speed that is faster than the motion of the atoms in a quantum gas (few mm/s) is necessary to make sure that the electron beam effectively interacts with an unperturbed cloud. Given a full transmission of the ions to the detector and assuming a well adjusted ion detector with 50$\%$ overall detection efficiency, about 20$\%$ of the atoms can be detected. This signal is enough to extract the accurate density distribution from a quantum gas but puts limitations on the ability to count atoms.

In scanning electron microscopy, multiple scattering and secondary scattering processes often pose a problem for the proper interpretation of the signal. As the dilute atomic cloud collides only with a fraction of 10$^{-5}$ of the incident electrons, multiple scattering is completely negligible. Concerning secondary processes however, ion-atom collisions are of some importance. As the ions are created inside the gas, they can scatter with other atoms on their way out of the gas. As the potential between an ion and an atom scales as $r^{-4}$, where $r$ is the relative distance, the scattering cross section become as a large as $10^{-12}$\,cm$^2$ when low temperatures are approached~\cite{Wurtz_2010, Cote_2000}. For higher densities we observe indeed a non-exponential atom loss which indicates the presence of secondary collisions~\cite{Wurtz_2010}. Depending on the specific imaging mode and investigation these processes have to be taken into account. They can be suppressed by reducing the transverse extension of the atomic gas or by increasing the electrostatic extraction field as the cross section rapidly drops with the ion energy~\cite{Cote_2000}. 

\subsection{Experimental setup}

The experiment combines an apparatus for the production of $^{87}$Rb Bose-Einstein condensates with the setup of a scanning electron microscope. We limit the presence of magnetic fields that can distort the electron beam by making an all-optical BEC~\cite{Gericke_2007} and by shielding the main chamber with $\mu$ -metal. Except for these additional cares an ion detection path is implemented and a test target can be moved into the electron beam for calibration purposes. A detailed description of the electron column, the alignment and characterization of the electron beam, the ion optics and the detection path is given below.

\subsubsection*{The electron column.}

\begin{figure}[t!]
\begin{center}
\includegraphics[width=0.7\textwidth]{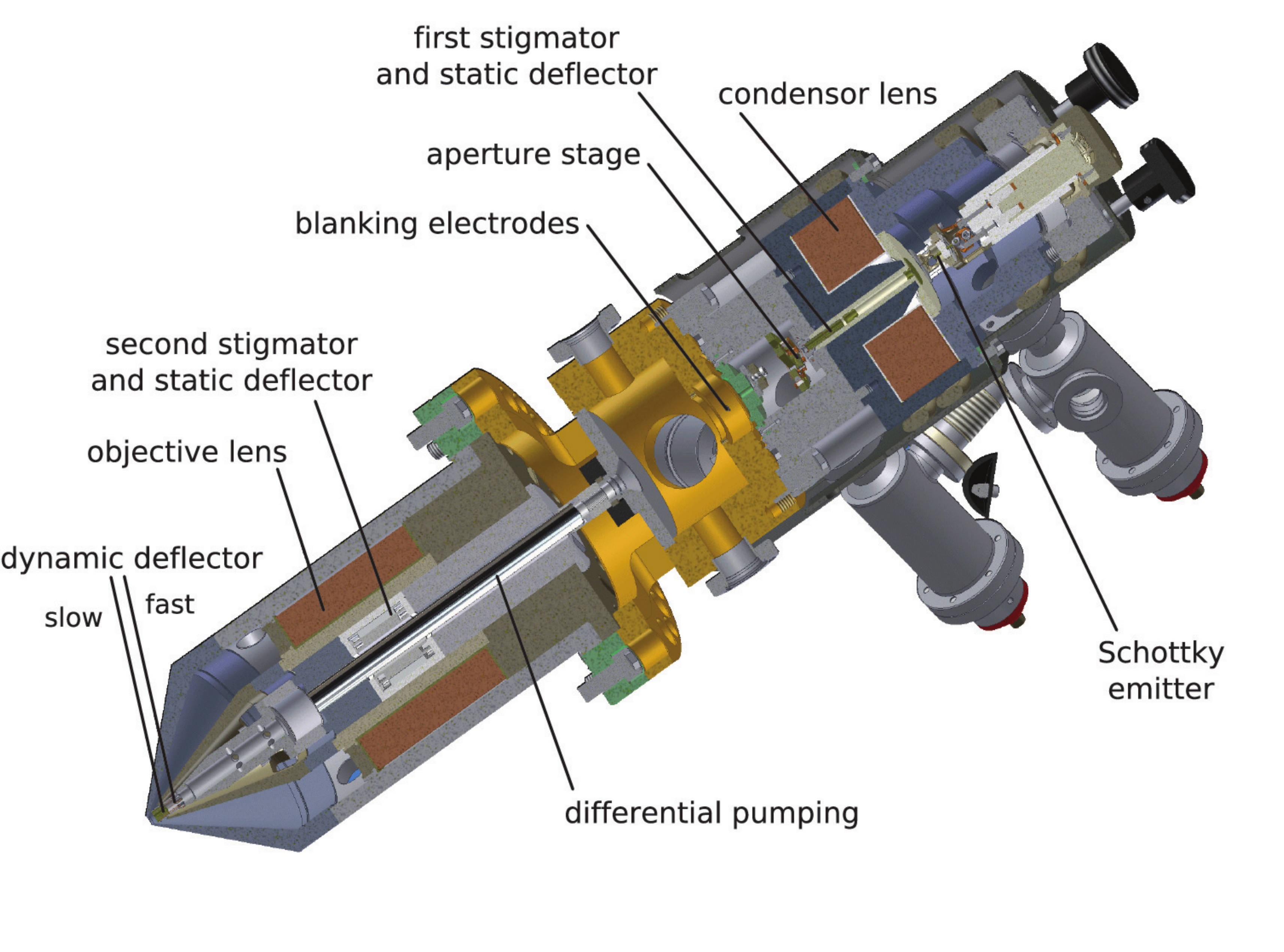}
\end{center}
\caption{Technical drawing of the electron column showing various parts required for obtaining a highly focused electron beam for the experiment.} 
\label{column} 
\end{figure} 

The electron column is mounted on top of the main chamber. It is a custom made column with thermal ZrO-Schottky emitter, which provides an electron beam with an energy up to 6 keV ~\cite{Gericke_2010}. The upper part of the column consists of three different vacuum chambers: the gun chamber, which holds the electron emitter, the aperture chamber, in which a movable stage with several apertures is mounted, and the so called intermediate chamber, which consists of a pneumatic isolation valve connected to the main vacuum chamber. This valve is open during normal operation, and closed when it is necessary to isolate the main chamber from the electron column when this one is vented to substitute the emitter (12000 hours lifetime). The lower part of the column, consisting of a pole piece made of an iron-nickel alloy extends into the main chamber (see Fig.~\ref{column}). The current of the electron beam is measured with a Faraday cup, which is placed 5 cm underneath the tip. The electron column has two magnetic lenses to focus the beam at a working distance of 13 mm below the tip. The magnetic field produced inside the second lens, which is close to the atoms, reaches 2000 G. An efficient shielding provided by the $\mu$ -metal pole piece of the tip reduces the field to 1 G at the position of the atoms. The effect of spherical aberrations is corrected by inserting apertures into the electron beam. Astigmatic aberrations and displacements of the electron beam, caused by the lenses, are corrected by electrostatic stigmators and deflectors, which are placed in a combined unit behind each lens~\cite{Hawkes_1996}. The movable aperture strip with apertures of diameter of 300 $\mu$m, 250 $\mu$m, 150 $\mu$m, 100 $\mu$m, 50 $\mu$m and 20 $\mu$m is placed underneath the first deflector and stigmator stage in the aperture chamber together with an electrostatic blanking unit that allows for the dumping of the electron beam inside the column with a frequency up to 5 MHz. Two physically separated scanning units, consisting of electrostatic quadrupoles, move the electron beam over the atomic cloud. The fast scan unit has a field of view of 200 $\mu$m $\times$ 200 $\mu$m and can be moved with a bandwidth of 10 MHz. The slow scan unit provides scans up to 20 kHz and a field of view of 1 mm $\times$ 1 mm.

\subsubsection*{Alignment and characterization.}

For the alignment and the characterization of the electron beam we use two different test targets mounted on a movable holder. The first target is a copper mesh and the second is a hole with a diameter of 200 $\mu$m. A conducting plate is placed on the holder to detect secondary and backscattered electrons from the test targets. The alignment consists of adjusting the two lenses, deflectors and stigmators to maximize the resolution of the image in a convenient field of view. For a coarse alignment the mesh target is used and a follow up fine adjustment is performed on the hole target. This is also used to determine the beam diameter by taking a set of line scans over the edge of the hole. Each of the measured scans is fitted with an error function, from which we derive the beam waist. Finally, we remove both targets and optimize the alignment on the atoms in a deep 2D optical lattice with spacing of 600 nm~\cite{Gericke_2010,Wurtz_2008,Wurtz_PRL_2009}. The highest spatial resolution which we can obtain for a typical current of 20 nA is $\sim$ 100 nm. The depth of focus is derived by measuring the FWHM of the electron beam at different vertical positions without refocusing. This is achieved by moving the hole target along the beam axis by a micrometer positioning translation stage. The Rayleigh length is obtained as 35 $\mu$m assuming a Gaussian shape of the electron beam with a waist of $\sim$ 100
nm. The vertical extension of the atomic cloud is about 6 $\mu$m such that a depth of focus of 35 $\mu$m guaranties a constant electron beam diameter across the atomic sample.

\subsubsection*{Ion optics and detection.}

Once the atoms are ionized, they are guided by a series of electrodes to a channeltron detector. The ions first hit a conversion electrode ($-4.8\,$kV), where secondary electrons are produced and further accelerated towards the channeltron. The input of the channeltron has a voltage of -2.2\,kV. A detected ion produces a negative voltage peak of -10 to -60\,mV with a pulse width of about 100\,ns. This signal is converted into a TTL pulse by a discriminator and amplifier. The pulses are recorded with 10\,ns time resolution by the control hardware (ADwin-Pro II).

\subsection{Imaging ultracold quantum gases}

\begin{figure}[t!]
\begin{center}
\includegraphics[width=0.7\textwidth]{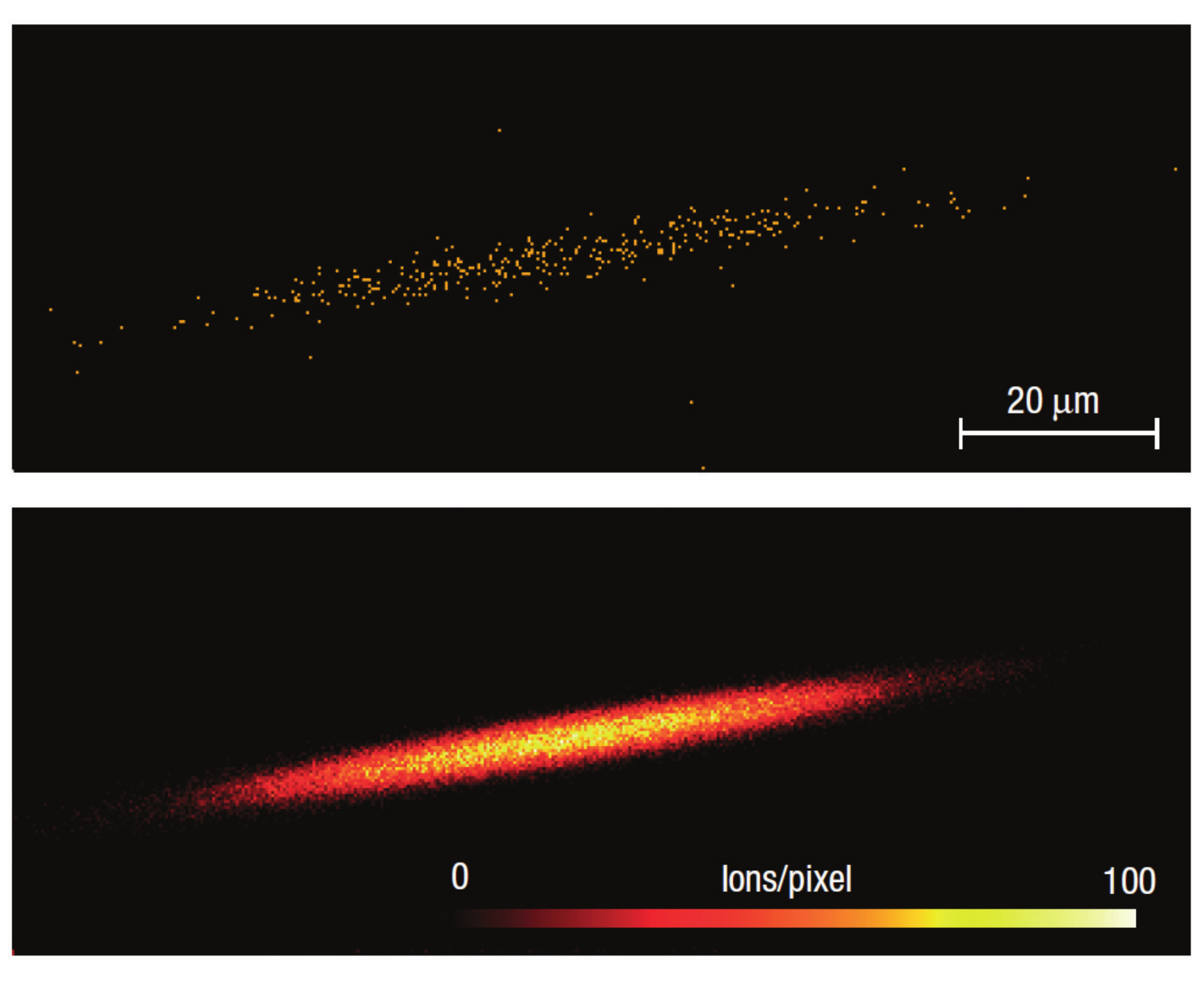}
\end{center}
\caption{Top: Image of a Bose-Einstein condensate of 10$^5$ $^{87}$Rb atoms. The image has 400 $\times$ 150 pixels with a pixel size of 300 nm $\times$ 300 nm. Each pixel was illuminated for 2 $\mu$s with the electron beam (140 nm FWHM beam diameter). Every dot corresponds to a detected atom. In total, 350 ions were collected during the exposure. Bottom: Sum over 300 images. Each image was taken in a separate experimental run. From Ref.~\cite{Gericke_2008}.} 
\label{BEC_image} 
\end{figure} 

A typical image of a Bose-Einstein condensate is shown in Fig.~\ref{BEC_image}. Summing over many individual images taken in different experimental runs yields a high precision, high contrast density profile of the condensate. From the given beam parameters, it is obvious that the spatial resolution of a SEM exceeds the diffraction limit of an optical microscope. However, an improvement of the resolution is achieved at the cost of a smaller beam current. As all characteristic length scales in a quantum gas are in the order of a few hundred nanometer (the healing length is $\sim$ 400 nm, the typical interatomic distance is $\sim$ 200 nm, the typical spacing in an optical lattice is $\sim$ 500 nm, a resolution of 100\,nm is sufficient for most purposes. For this reason, we run the electron column at working points which are rather unusual for scanning electron microscopes and are optimized for the experimental requirements. Typical combinations of beam current and beam diameter that are used in the setup are summarized in Table~1.

\begin{table}[h]
\begin{center}
		\begin{tabular}{| c | c|}
%		\begin{tabular}{|l|r|}
		\hline
		beam diameter (FWHM)	& beam current	\\\hline
		90\,nm													&12\,nA										\\
		250\,nm													&100\,nA									\\
		400\,nm													&180\,nA								\\
		5000\,nm												&800\,nA							\\\hline
		\end{tabular}
		\caption{Electron beam characteristics for some of the typical settings in the experiment.}
		\end{center}
\label{table1}
\end{table}

\subsection{Single site addressing}

\begin{figure}[t!]
\begin{center}
\includegraphics[width=0.7\textwidth]{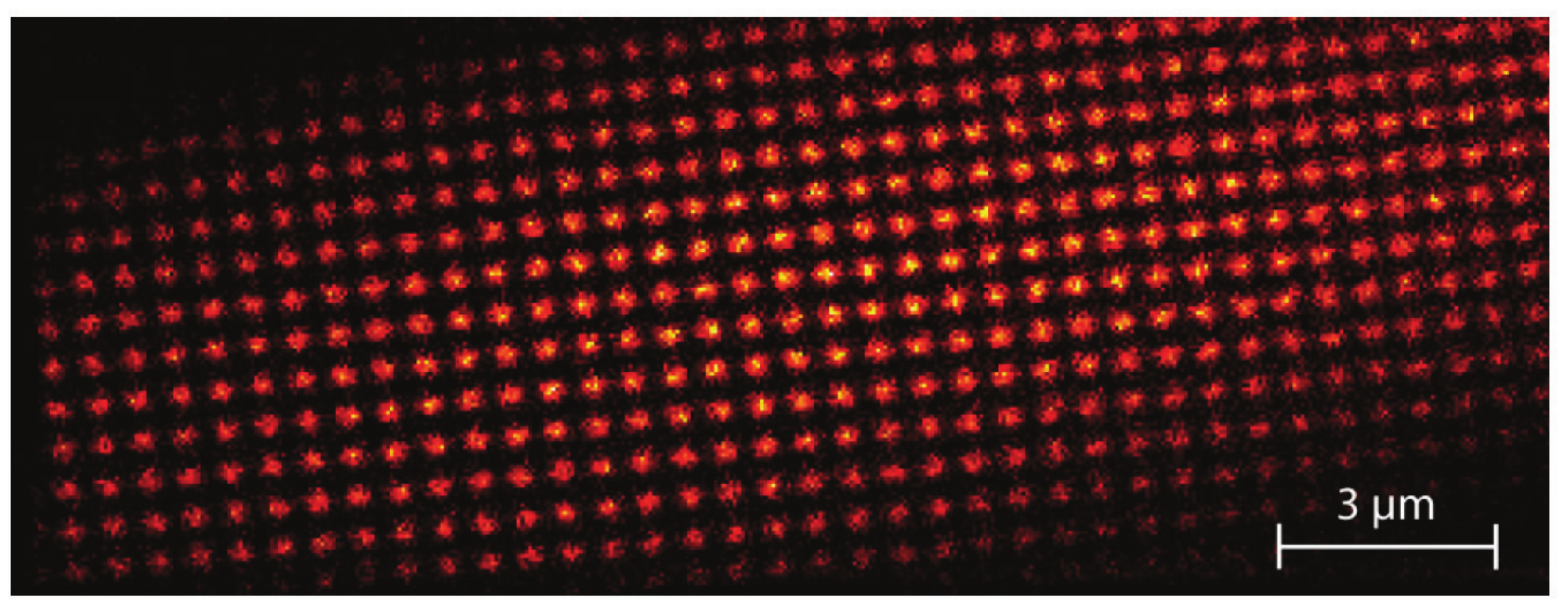}
\end{center}
\caption{Image of a Bose Einstein condensate loaded in a two-dimensional optical lattice with 600 nm lattice spacing (sum obtained from 260 individual images). Each site has a tube-like shape with an extension of 6 $\mu$m perpendicular to the plane of projection. The central lattice sites contain about 80 atoms. From Ref.~\cite{Wurtz_PRL_2009}.} 
\label{lattice2d} 
\end{figure} 

\begin{figure}[t!]
\begin{center}
\includegraphics[width=0.5\textwidth]{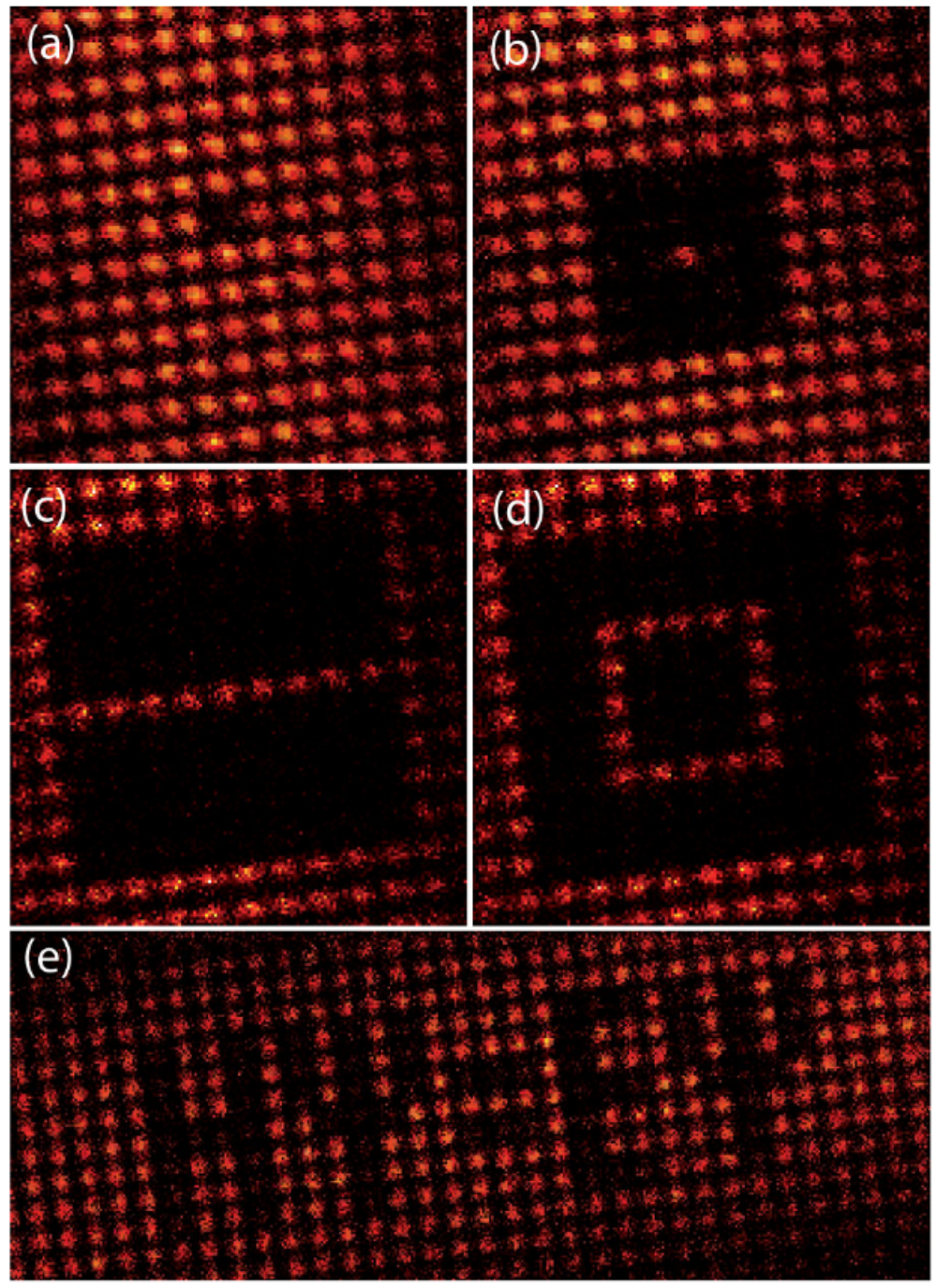}
\end{center}
\caption{Engineering arbitrary patterns of atoms loaded in a 2D optical lattice with a spacing of 600 nm. Each emptied site was illuminated with the electron beam between 1 and 3 milliseconds before the imaging procedure was started. From Ref.~\cite{Wurtz_PRL_2009}.} 
\label{address2d} 
\end{figure} 

To characterize the resolution of the imaging technique, we load the condensate in a two-dimensional optical lattice with 600\,nm lattice period (Fig.~\ref{lattice2d}). The structure of the potential is clearly visible and from a quantitative evaluation we can deduce a spatial resolution of better than 150\,nm~\cite{Gericke_2007}.
With such a high resolution, the technique can immediately be used for single-site addressing in an optical lattice. To this purpose, we selectively remove atoms from individual sites by means of collisions with the focused electron beam. In this way, arbitrary patterns of occupied lattice sites can be produced~\cite{Wurtz_PRL_2009}. Four elementary examples are presented in Fig.~\ref{address2d}. A single defect in the lattice structure is shown in the first panel and represents an ideal starting point to study the tunneling dynamics at a single lattice site. The opposite situation corresponds to an isolated lattice site and is shown in the second panel. Such a mesoscopic ensemble provides, for instance, the possibility to study the transition from few-body systems to the thermodynamic limit. It can also act as a paradigm for Rydberg blockade studies as the spatial extension of the ensemble is very small. Self trapping and many-body tunneling processes can also be studied with such an initial density distribution. A chain and a ring of lattice sites are shown in the lower panels in order to illustrate the large variety of achievable geometries. The approach allows for any arbitrary pattern that fits to the underlying quadratic geometry (Fig.~\ref{address2d}(e)). Upon illumination, the number of remaining atoms in a site decays exponentially in time. In terms of writing time, a lattice site with less than 5 \,\% filling can be prepared in about 1\,ms, being sufficient to extensively tailor and structure the quantum gas in the optical lattice. 

\section{Experiments}

In this section, we present three examples how quantum gases can be investigated by means of SEM. For weak probing (beam current $<$ 50\,nA), the electron beam acts as a small perturbation to the system and density distributions and pair correlation functions can be measured. For strong probing (beam current $>$ 50\,nA), the back action on the atoms becomes important and we observe quantum Zeno dynamics in the system. In the third example, we show how the excitation to Rydberg states can even further enhance the interaction between the electron beam and a Bose-Einstein condensate.

\subsection{Detection of pair correlations in temporal domain}

The measurement of correlations is an important diagnostic tool to characterize an interacting many-body quantum gas. First order correlations are often studied via interference experiments. Second order correlations can be determined, e.g., by photoassociation~\cite{Kinoshita_2005}, by the detection of intensity noise \cite{Foelling_2005} or by a single particle sensitive detection technique. The second order correlation function $g^{2}(\tau)$ represents the conditional likelihood for detecting a particle at a time $\tau$ after a previously detected particle and quantifies second order coherence. A BEC can be represented as a single macroscopic coherent wave function. Hence, for a pure BEC one expects $g^{2}(0)=1$. Well above the BEC critical temperature the thermal atoms behave like incoherent particles. The bosonic nature of the atoms leads to bunching, $g^{2}(0)=2$. Using scanning electron microscopy as a time-resolved local detection method, we can directly measure second and even third order temporal correlation functions~\cite{Guarrera_correlations_2011}. To this end, the electron beam is pointed continuously in the center of the trapped gas for a given time and the time-resolved ion signal is analyzed.

The general form of the normalized spatio-temporal correlation function of $n$ particles at position $\textbf{r}_i$ at time $t_i$, with $i=1,...,n$, is given by:

\begin{equation}
g^{(n)}(\textbf{r}_1,t_1;...;\textbf{r}_n,t_n)= 
\frac{\langle \widehat{\Psi}^\dagger(\textbf{r}_1,t_1)... \widehat{\Psi}^\dagger(\textbf{r}_n,t_n) \widehat{\Psi}(\textbf{r}_n,t_n)...  \widehat{\Psi}(\textbf{r}_1,t_1) \rangle}{\langle \widehat{\Psi}^\dagger(\textbf{r}_1,t_1) \widehat{\Psi}(\textbf{r}_1,t_1)\rangle...\langle \widehat{\Psi}^\dagger(\textbf{r}_n,t_n)\widehat{\Psi}(\textbf{r}_n,t_n)\rangle}
\label{eq:g2 general}
\end{equation}
where $\widehat{\Psi}$ and $\widehat{\Psi}^\dagger$ are the bosonic field operators and $\langle ... \rangle$ indicates the ensemble average. We first consider a thermal gas. An analytical expression for $g^{(1)}(\textbf{r}_1,t_1;\textbf{r}_2,t_2)$ can be derived, extending the approach of Ref.~\cite{Naraschewski_1999} to the temporal domain. For an ideal Bose gas at temperature $T$ above the critical temperature $T_c$, trapped in a harmonic potential with average trapping frequency $\omega$, we obtain:
\begin{equation}
g^{(1)}(r,\tau)=\frac{1}{\left(1+i \frac{\tau}{\tau_c} \right)^{3/2}}\exp \left( -\frac{m r^2}{2\hbar \tau_c^2}\frac{\tau_c+i\tau}{1+\left(\frac{\tau}{\tau_c}\right)^2}\right)
\label{eq:g1}
\end{equation}
where $\tau_c=\frac{\hbar}{k_B T}$ is the correlation time, $\tau=t_2-t_1$ and $r=|\textbf{r}_2-\textbf{r}_1|$. The above result is valid in the limit $\omega\tau,  \hbar\omega/(k_BT)\ll1$. From Eq.\,\ref{eq:g1}, we can derive any higher order correlation function for thermal bosons using Wick's theorem. In particular, the second order correlation can be calculated as $g^{(2)}(r,\tau)=1+\vert g^{(1)}(r,\tau) \vert^2 $.

\begin{figure}[t!]
\begin{center}
\includegraphics[width=0.7\textwidth]{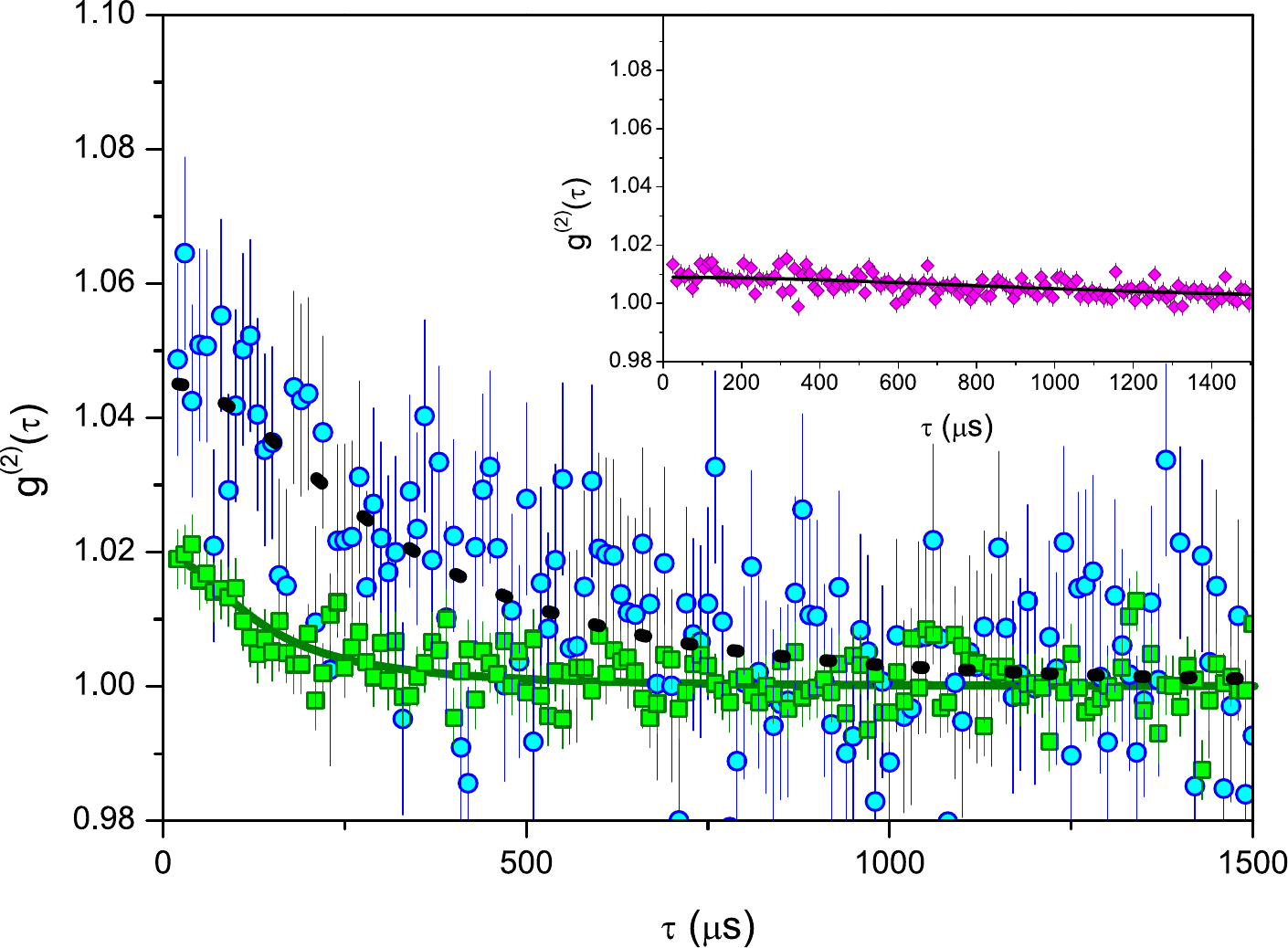}
\end{center}
\caption{Temporal second order correlation function. Data for two different temperatures (circles for $45$ nK, squares for $100$ nK) are plotted together with the theoretical prediction. The inset shows the pair correlation function of a BEC. Please note that even well below the critical temperature (the thermal fraction cannot be detected in
time-of-flight absorption imaging) we are able to measure a small residual bunching induced by the thermal component. From Ref.~\cite{Guarrera_correlations_2011}.} 
\label{fig:g2}
\end{figure}

\begin{figure}[t!]
\begin{center}
\includegraphics[width=0.7\textwidth]{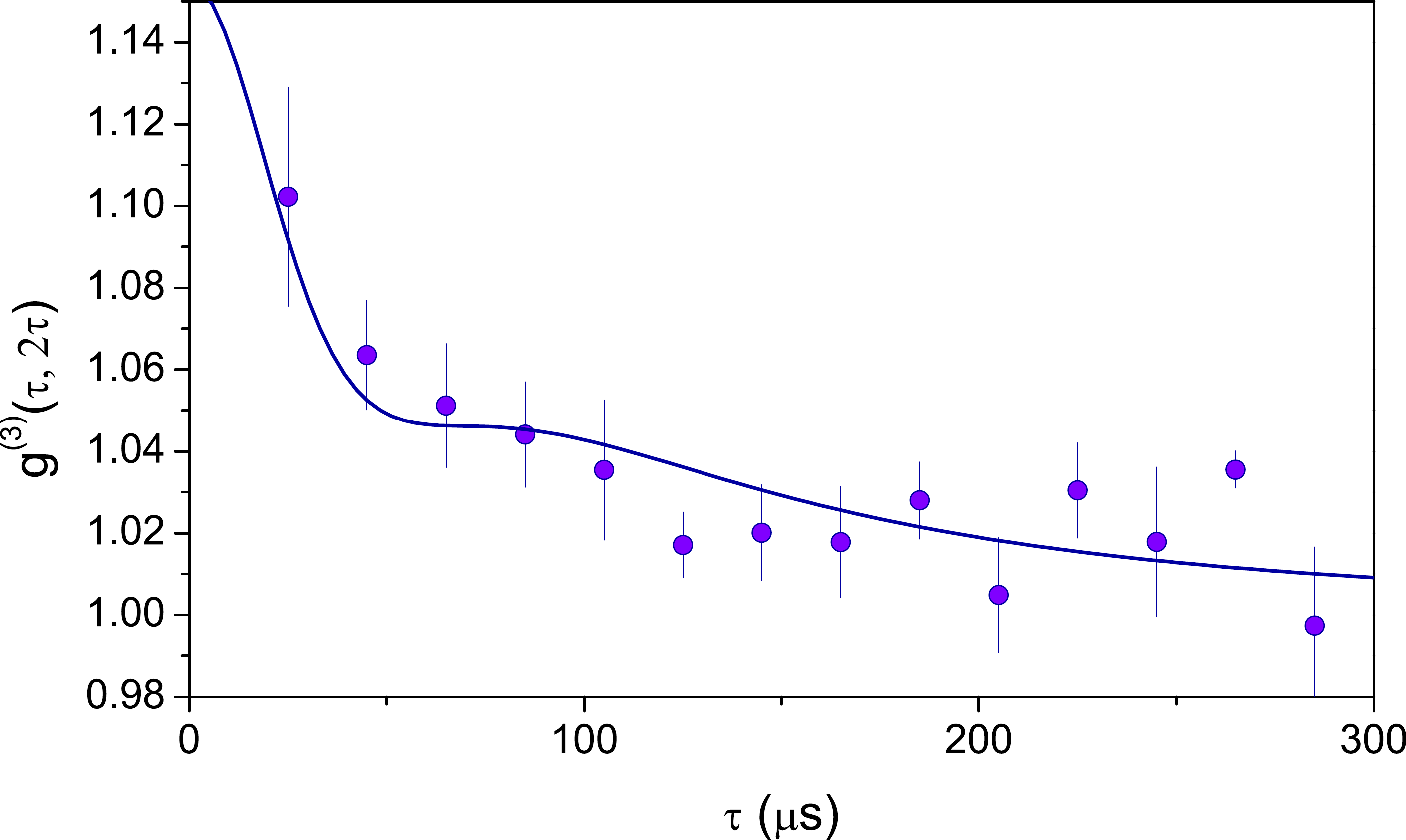}
\end{center}
\caption{Temporal third order correlation function for a thermal gas. The data (dots) shows the result for a thermal cloud with $T=100$\,nK. The signal has been integrated along one time-axis ($\tau_1=t_2-t_1,\tau_2=t_3-t_1=2\tau_1$). The solid line is the theoretical prediction. From Ref.~\cite{Guarrera_correlations_2011}.} \label{fig:g3} 
\end{figure}

In Fig.~\ref{fig:g2} we show the measured $g^{(2)}(0,\tau):=g^{(2)}(\tau)$ of a thermal cloud of atoms for two different temperatures $T=45$\,nK and $T=100$\,nK. The finite extension of the cloud along the electron beam axis leads to a reduction of the contrast as the signal is integrated over many phase space cells. This can be accounted for by convolving the pair correlation function with the density profile of the trapped gas \cite{Guarrera_correlations_2011}. Fluctuations in the total number of detected ions affect the normalization of the correlation function, which is obtained by averaging over several experimental cycles. This causes an offset shift of the uncorrelated signal to a value $1\%$ above $1$. In order to compensate these fluctuations we normalize $g^{(2)}(\tau)$ by the factor $1+\sigma^2/ \langle N \rangle ^2$, where $\sigma^2$ and $\langle N \rangle$ are respectively the variance and the mean value of the total number of detected ions in the different experimental realizations. Experiment and theory are in good agreement and the different correlation times are clearly visible. A measurement with a BEC is shown in the inset of Fig.~\ref{fig:g2} and shows no noticable correlations.

The measurement scheme can also be extended to evaluate third order correlation functions. The results are shown in Fig.~\ref{fig:g3}. The solid line is the theoretical prediction based on Eq.\,\ref{eq:g1},
$g^{(3)}(\tau_1,\tau_2)=1+\vert g^{(1)}(\tau_1)\vert^2 + \vert g^{(1)}(\tau_2)\vert^2 +\vert g^{(1)}(\tau_2 -\tau_1)\vert^2+2Re(g^{(1)}(\tau_1)g^{(1)}(\tau_2)g^{(1)}(\tau_2-\tau_1))$, with $\tau_1=t_2-t_1$ and $\tau_2=t_3-t_1$, leaving the global amplitude as the only free parameter. As a consequence of the factorial law $n!$ bunching is expected to be more pronounced at higher orders $n$. For this reason higher order correlations can be employed as a highly sensitive test for coherence, with the only drawback represented by the need of high statistics. 

Having confirmed the suitability of SEM for the detection of pair correlations, we move on to a system with strong quantum correlations: a one-dimensional (1D) Bose gas~\cite{Guarrera_2012}. We prepare an array of 1D Bose gases with help of a blue-detuned two-dimensional optical lattice in which we load a BEC. For sufficiently deep lattices, the tubes are completely decoupled and the effective interaction strength within each tube is given by $g_{\mathrm 1D}\simeq 2a\hbar\omega_r$, where $a$ is the $s$-wave scattering length and $\omega_r$ the frequency of the radial harmonic confinement. The strength of the correlations within the 1D gas can be expressed in terms of the Lieb-Liniger parameter $\gamma(x)=mg_{\mathrm 1D}/\hbar^2 \rho(x)$, which in local density approximation becomes position dependent ($x=0$ corresponds to the center of the tube). Here, $m$ is the mass of the particle and $\rho(x)$ is the line density. For $\gamma(0)\ll1$ and $T$ below the degeneracy temperature $T_d=\hbar^2\rho^2(0)/2mk_B$~\cite{Kheruntsyan_2005}, a weakly interacting quasi-condensate phase is formed. For such a phase, the spatial density profile in a harmonic trap is expected to be well described by a Thomas-Fermi (TF) parabola~\cite{Dunjko_2001} and $g^{(2)}(0,0)\approx1$~\cite{Kheruntsyan_2005}. In the opposite limit where $\gamma(0)\gg1$, the 1D gas is strongly interacting and can be approximated as a Tonks-Girardeau (TG) gas~\cite{Kinoshita_2004, Paredes_2004}. For such a gas the density profile is a square root of a parabola and $g^{(2)}(0,0)\ll1$, which indicates an effective reduction in the overlap of the particle wavefunctions, resembling the fermionic exclusion principle. As a consequence, the local pair correlation function is strongly reduced \cite{Kinoshita_2005}.

\begin{figure}[t!]
\begin{center}
\includegraphics[trim={0.3cm 6.8cm 0cm 6.1cm},clip,width=0.9\textwidth]{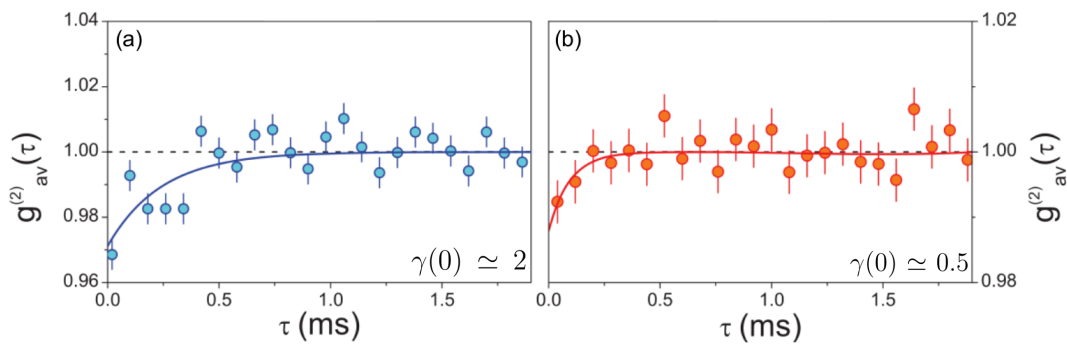}
\end{center}
\caption{Antibunching in a strongly correlated 1D Bose gas. Pair correlation function for $1\times10^4$ (a) and $6\times10^4$ (b) trapped atoms, representing two different 1D interaction strengths. The solid lines are an exponential fit to the data. From Ref.~\cite{Guarrera_2012}.} \label{fig:correlation}
\end{figure}

We can experimentally access both regimes by varying the density and the radial confinement $\omega_r$. We perform measurements for $\gamma(0)\simeq 2$, representing the onset of strong interactions and $\gamma(0)\simeq0.5$, representing weaker interactions. We measure the pair correlation function in the same way as described above. The results are shown in Fig.\,~\ref{fig:correlation}. In both cases the antibunching that is due to the repulsive interaction between the atoms is visible. The amplitude is only a few percent, as the signal is integrated over 8-10 independent tubes. Note that the time-dependent pair correlation function for a strongly interacting 1D Bose gas cannot be mapped on a corresponding fermionic system, because the annihilation operators at two different times do not commute.

Pair correlations are a sensitive measure for the dynamics of a many-body quantum system. We have demonstrated that a local probe such as SEM is well suited to measure them. Care has to be taken regarding the prospective measurement time as the low detection efficiency enters quadratically in a pair correlation measurement.

\subsection{Dissipation induced quantum Zeno effect}

When the intensity of the electron beam exceeds the weak probing limit, the back action on the quantum gas becomes important. Then the electron beam (EB) takes over the role of an environment to which the quantum gas is coupled. In the context of the theory of open quantum systems, environmental action gives rise to effective Hamiltonians that can contain imaginary terms~\cite{Omnes_1992,Wiseman_2010,Barreiro_2011}. The losses induced by the electron beam can be regarded as a dissipation process leading to such an imaginary potential. As a result, a generalized version of the quantum Zeno effect occurs. The quantum Zeno effect can be described as the inhibition of the decay of any unstable quantum state by sufficiently frequent measurement of that state~\cite{Misra_1977,Kofman_2000}. If the rate of measurement is high enough, the quantum dynamics is slowed down~\cite{Itano_1990}. Here, it is the electron beam that measures the position of the atoms in the BEC.

In the experiment, we locally remove atoms from a Bose-Einstein condensate, changing the intensity of the electron beam \cite{Barontini_2013}. The condensate can be described to a good approximation by a mean-field wavefunction obeying the Gross-Pitaevskii equation (GPE). This remains valid even if the condensate is coupled with the environment. Starting from the Lindblad master equation $i\hbar \partial_t\hat{\rho} = [\hat{H},\hat{\rho}]+ i\hbar\hat{\mathcal{L}}\hat{\rho}$, where $\hat{\rho}$ is the density operator of the many-body system, $\hat{H}$ is the Hamiltonian operator, and $\hat{\mathcal{L}}$ is the dissipation operator such that $\hat{\mathcal{L}}\hat{\rho}=-\int d\textbf{x}\gamma(\textbf{x})/2
[\hat{\Psi}^+\hat{\Psi}\hat{\rho}+\hat{\rho}\hat{\Psi}^+\hat{\Psi}-2\hat{\Psi}\hat{\rho}\hat{\Psi}^+]$, with $\gamma(\textbf{x})$ the local dissipation rate, we can write the equation of motion for the expectation value of the bosonic field operator $\hat{\Psi}$ as $i\hbar\partial_t\langle\hat{\Psi}\rangle=Tr(\hat{\Psi}\partial_t\hat{\rho})$, which leads to a time-dependent GPE with an additional imaginary term \cite{Brazhnyi_2009,Witthaut_2011}:
\begin{equation}
i\hbar\frac{\partial\psi(\textbf{x},t)}{\partial t}=\left(-\frac{\hbar^2\nabla^2}{2m}+V_{ext}+g|\psi(\textbf{x},t)|^2-i\hbar\frac{\gamma(\textbf{x})}{2}\right)\psi(\textbf{x},t).
\label{dGPE}
\end{equation} 
Here $\psi$ is the BEC wavefunction, obeying the constraint $\int|\psi(\textbf{x},t)|^2d\textbf{x}=N(t)$, $V_{ext}$ the trapping potential and $g=4\pi\hbar^2a/m$, $a$ being the s-wave scattering length. The technique allows the independent control of the Hamiltonian and the dissipative terms.

\begin{figure}[t!]
\begin{center}
\includegraphics[width=9cm]{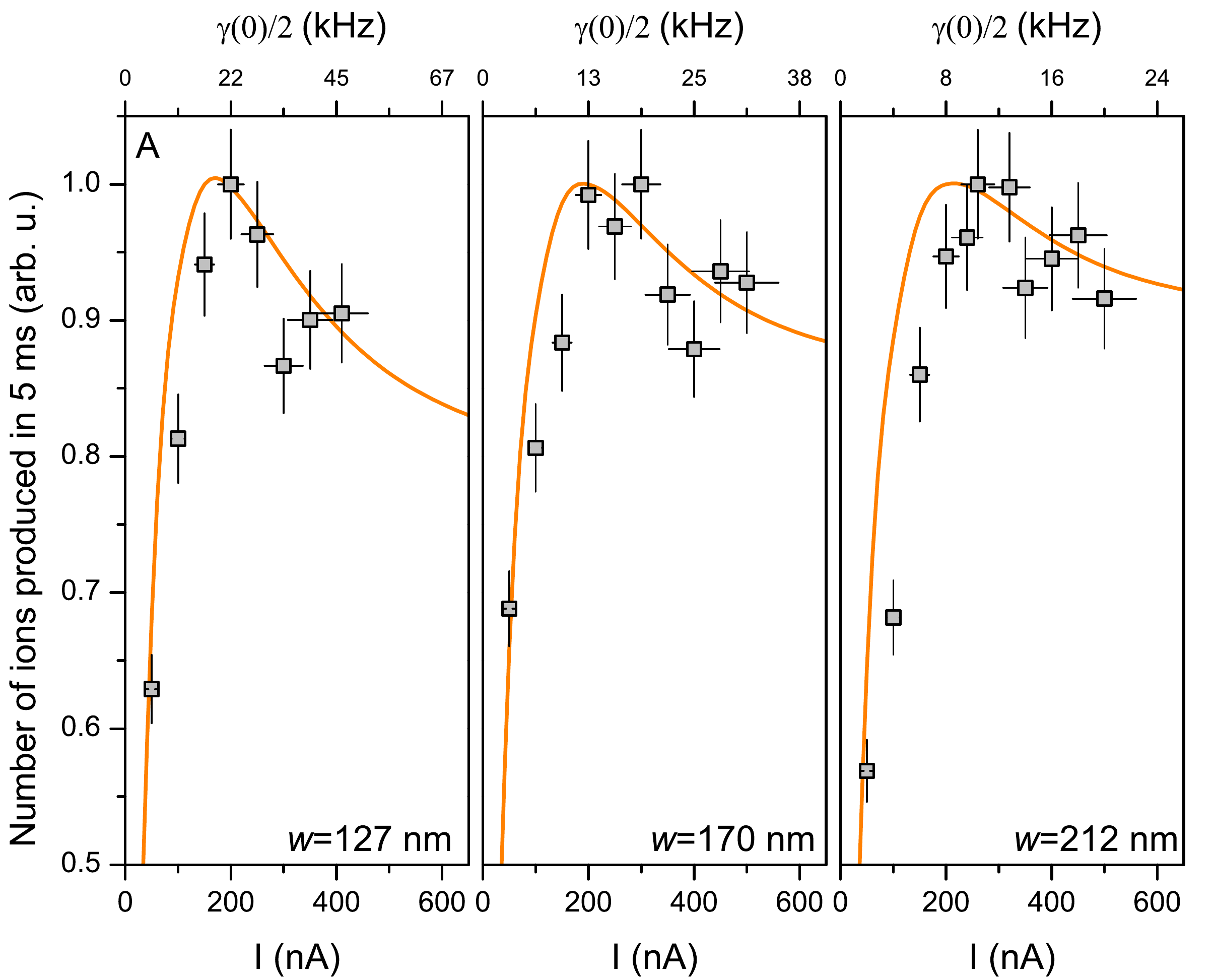} 
\end{center}
\caption{Quantum-Zeno dynamics in a Bose-Einstein condensate. Number of ions collected within the first 5 ms of continuous dissipation on a BEC as a function of the EB current $I$. The three panels report the data obtained with $w= 127 (5), 170 (7)$ and $212 (8)$\,nm, from left to right. Each data point is the average over 75 experimental repetitions. The solid lines are the result from numerical simulations. From Ref.~\cite{Barontini_2013}.}
\label{Zeno_BEC}
\end{figure}

In the experiment, the position of the EB is fixed at the center of the BEC and we record the temporal signal from the ion detector. By controlling the beam parameters, we can engineer the dissipative term in Eq. \ref{dGPE} as $\gamma(\textbf{x})=I\sigma/(2\pi e w^2)\exp(-(x^2+y^2)/2w^2)$, where $I$ is the EB current, $\sigma$ the electron-atom scattering cross-section, $e$ the elementary charge and $w$ the standard deviation of the electron density distribution, assumed to be gaussian~\cite{Barontini_2013}. In Fig.~\ref{Zeno_BEC}, for three different values of $w$, we show the number of ions collected in the first 5 ms of continuous dissipation as a function of the EB current. We observe a non-monotonic dependence of the number of ions produced as a function of the EB current for all three dissipation strengths: above a critical value of the EB current, the signal decreases despite the stronger dissipation rate. This paradoxical behavior is more pronounced for smaller values of $w$ and is the signature of quantum Zeno dynamics. Results obtained by numerically solving Eq.~\ref{dGPE}, additionally taking into account secondary effects like ion-atom collisions \cite{Barontini_2013}, are compared in the same figure. The very good agreement  between the data and the numerically solved model demonstrates that the description of the EB as a pure dissipative potential is sufficient to capture the main features of the observed dynamics. A localized imaginary potential induces total reflection as the strength of the potential $U$ goes to infinity~\cite{Barontini_2013,Allcock_1969_1,Allcock_1969_2,Allcock_1969_3}. The EB current corresponding to the maximum ion production is of special importance, since it sets the parameters that make it possible to engineer the most efficient absorbing potential. At the same time, the maximum separates two dynamical regimes: for small dissipation strength, the unitary Hamiltonian part dominates, while for large dissipation strength, the unitary part dominates.

The results show that the controlled coupling of a quantum gas to an environment can lead to interesting dynamics and new ways to engineer many-body quantum states. Engineering dissipative processes has already proven to be useful for the generation of many-body quantum states \cite{Syassen_2008}. In the future, we expect that such concepts will be extended to create quantum states that are not reachable otherwise \cite{Zoller_review_2012}.

\subsection{Electron microscopy of a Rydberg excited Bose-Einstein condensate}

Rydberg atoms combine strong dipole-dipole interactions with long radiative lifetimes and are a promising candidate to implement long-range interactions~\cite{Schauss_2012}. The possibility to image Rydberg atoms gives direct access to spatial correlation functions and is the key component to study new quantum phases~\cite{Cinti_2010}, tailored spin systems~\cite{Lee_2011} and energy transfer mechanisms~\cite{Wuester_2011}. Direct Rydberg atom imaging has been demonstrated so far in optical lattices~\cite{Schauss_2012}, with ion emission microscopy techniques~\cite{Schwarzkopf_2011}, and with a scanning autoionization microscope~\cite{Lochead_2013}. Due to the huge electric transition dipole moment Rydberg atoms are much more sensitive to electric fields and feature huge cross sections for electron scattering. These are therefore well suited for studies in the context of SEM. Here, we show that the combination of scanning electron microscopy with Rydberg excited atomic samples offers new possibilities to prepare, image, and probe theses systems. 

\begin{figure}[t!]
\begin{center}
\includegraphics[width=16cm]{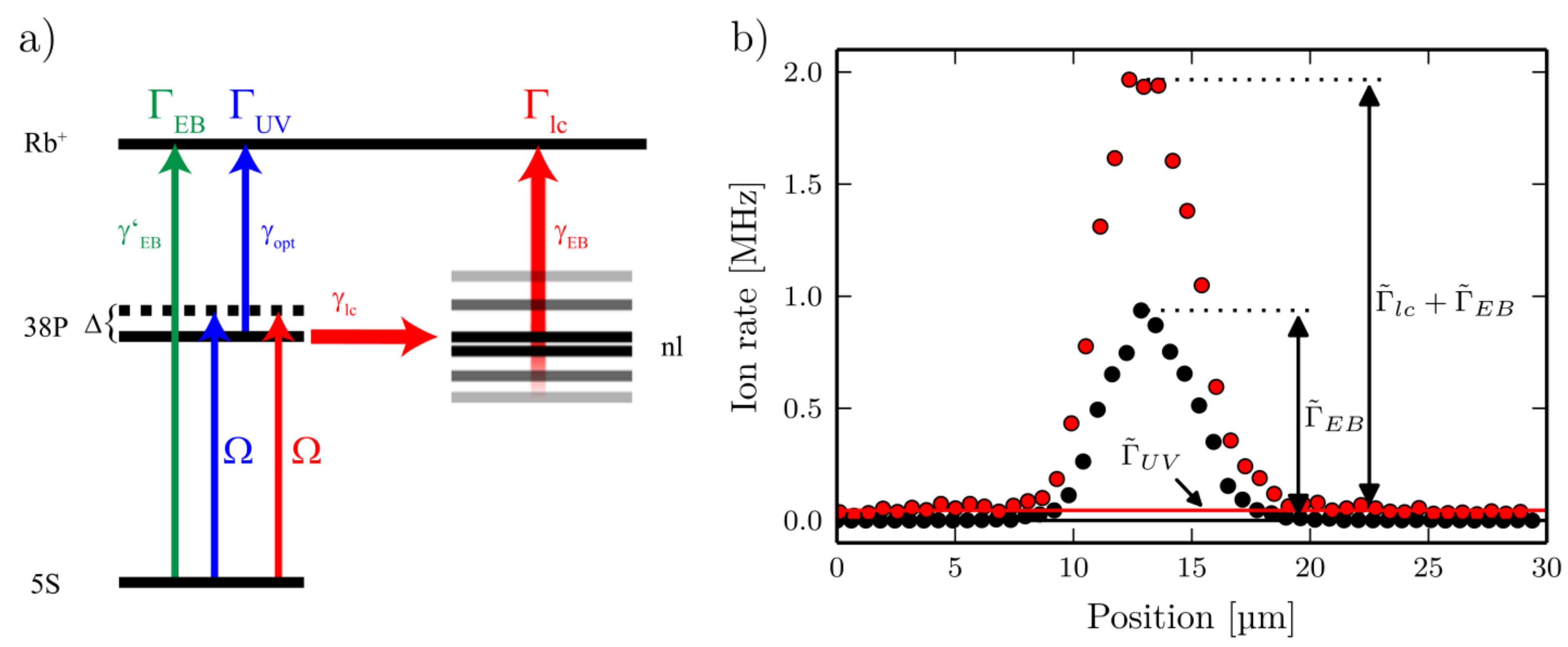} 
\end{center}
\caption{Scanning electron microscopy of a Rydberg excited Bose-Einstein condensate. \textbf{a}) Schematics of the three possible ionization channels for a single atom. The first process (effective ion rate $\Gamma_{EB}$, green) is direct electron impact ionization of ground state atoms. The second ionization channel (effective ion rate $\Gamma_{UV}$, blue) is a two step process, consisting of an optical excitation to a Rydberg state with Rabi frequency $\Omega$ followed by a photoionization process with rate $\gamma_{opt}$. The third ionization channel (effective ion rate $\Gamma_{lc}$, red) is a three step process that consists of an optical excitation to a Rydberg state with Rabi frequency $\Omega$, an $l$-changing collision (rate $\gamma_{lc}$) and electron impact ionization of the Rydberg states (rate $\gamma_{EB}$). \textbf{b)} Ion rates during an electron beam line scan (waist $w= 170(20)$ nm, current $I= 25(1)$ nA) over a BEC with $N_{tot}=150$k atoms. The red (black) dots show the signal with (without) the simultaneous irradiation of the UV laser ($I=7\,$mW, $w=700\,\mu$m, $\Delta=2\pi\times10\,$MHz). The given rates correspond to the processes described in a) - the tilde indicates that they denote the integrated signal coming from all atoms that contribute to the signal. From Ref.~\cite{Manthey_2014}.}
\label{Rydberg_SEM}
\end{figure}

We demonstrate the above mentioned possibilities with a Bose-Einstein condensate of rubidium ($N_{tot}=1.5\times 10^5$ atoms) that is off-resonantly excited to the $38P_{3/2}$ Rydberg state through a one-photon excitation by a high power ultra violet (UV) laser. Details of the experimental apparatus can be found in Ref.~\cite{Manthey_2014}. In order to investigate the effect of the single-photon Rydberg excitation on the performance of the electron microscopy, we perform continuous line scans over the BEC with and without the simultaneous illumination by the UV laser. The UV laser is blue-detuned ($\Delta=2\pi\times10$\,MHz) with respect to the $5S_{1/2}\rightarrow 38P_{3/2}$ transition. This ensures a long lifetime of the sample and avoids the excitation of molecular states consisting of a ground state atom and a Rydberg atom that appear on the red-detuned side of the resonance~\cite{Bendkowsky_2009, Butscher_2011}. Illuminating the sample with the electron beam and the UV laser, three different single-atom ionization channels are possible (see Fig.\,\ref{Rydberg_SEM}a): (i) direct electron impact ionization of ground state atoms, $\Gamma_{EB}$  (ii) photo-ionization by the trapping laser following an excitation by the UV laser, $\Gamma_{UV}$ and (iii) ionization originating from the combined action of the UV laser and the EB, $\Gamma_{lc}$. 

The effect of the individual processes is identified by analyzing the line-scans shown in Fig.\,\ref{Rydberg_SEM}b. The action of the UV light alone leads to a constant offset in the line scan of the Rydberg excited sample and the single-atom ion rate is measured to be $\Gamma_{UV}=0.3$ Hz. When the electron beam is scanned across the sample and no UV light is present, $\Gamma_{EB}$ can be directly measured. When both beams are switched on, the additional ion rate $\tilde\Gamma_{lc}$ appears. As can be seen from fig.~\ref{Rydberg_SEM}b this leads to an increase of the ion signal by a factor of 2. As the direct electron impact ionization of ground state atoms is unaltered by the UV light, the increase can only be explained by a much more efficient excitation of the Rydberg state by the UV laser. A quantitative analysis \cite{Manthey_2014} reveals a rate of $\Gamma_{lc}=1.4$ kHz, which is almost 3 orders of magnitude larger than the excitation rate of the UV light alone $\Gamma_{UV}=0.3$ Hz.

In order to understand this giant enhancement of the off-resonant excitation process in the presence of an electron beam we need to look at the interaction between an electron and a Rydberg atom. Inelastic collisions of the incoming electrons with the Rydberg atom lead to the population of nearby Rydberg states with different angular momentum $l$ and main quantum number $n$ (so-called $l$-changing collisions). The scattering cross section for these processes is huge and is given by~\cite{Omidvar_1973}
\begin{equation}
\sigma_{n,l}=\pi a_0^2\frac{MZ^2R}{m_eE}\left[A(n,l)\ln\left( \frac{m_e}{M}\frac{1}{Z}\frac{E}{R}\right)+B(n,l)\right]. 
\end{equation}
Here, $a_0$ is the Bohr radius, $m_e$ the electron mass, $M$ the mass of the atoms, $Z$ the charge of the nucleus, $E$ the collision energy and $R$ the ionization energy. $A(n,l)$ and $B(n,l)$ are numerical factors that depend on the quantum numbers $n$ and $l$ ~\cite{Omidvar_1973}. The collisions can be treated in first Born approximation for high impact energies, as present in the experiment. We calculate the $l$-changing cross section to be $\sigma_{n,l} (n=38,l=1)= 8.7\times 10^{-13}$ cm$^2$ for an EB energy of $6$\,keV. This cross section is orders of magnitude larger than the usual electron scattering cross section for ground state atoms. The resulting scattering rate into nearby Rydberg states is then given by
\begin{equation}
\gamma_{lc} = \sigma_l j /e ,
\end{equation}
where $e$ is the electron charge and $j=30$ A/cm$^2$ the current density of the electron beam. We obtain a scattering rate of $\gamma_{lc} = 150$\,MHz, which is much larger than any other decay or transition rate of the atom. The dynamics of the excitation process is therefore completely dominated by $l$-changing collisions. This can be cast into an effective linewidth given by $\gamma_{lc}$. As the linewidth exceeds the detuning $\Delta$, the optical excitation to the Rydberg state is very efficient. The $l$-changing cross section is two orders of magnitude larger than the electron impact ionization cross section for the Rydberg atom, which is given by $\sigma_{ion} = 4 \times 10^{-15}$ cm$^2$~\cite{Wurtz_2010, Vrinceanu_2005}. Once an atom is excited to a Rydberg state, it is scattered several hundred times in neighboring Rydberg states before it is eventually ionized. All this happens on a timescale below one microsecond, which means that every atom that is excited to a Rydberg state is eventually ionized. All three processes together explain the huge enhancement \cite{Manthey_2014}. 

The results show that the interaction of electrons with ultracold Rydberg atoms lead to drastic effects. The large ionization cross section make such a technique promising for the high resolution spatial detection of Rydberg atoms in atomic samples.

\section{Conclusions and future perspectives}

Scanning electron microscopy is a powerful technique for the study of microscopic structures and dynamical processes in ultracold quantum gases. It combines superb imaging capabilities with novel ways of system preparation. The ability of $in$ $situ$ and $in$ $vivo$ probing makes it well suited for the study of local and global non-equilibrium dynamics. The finite detection efficiency puts certain practical constraints on atom counting applications, but does not constitute a fundamental limitation. Compared to state-of-the-art high-resolution optical imaging \cite{Bakr_2009,Sherson_2010}, which features almost hundred percent detection efficiency, scanning electron microscopy provides a number of complementary features: (i) the \textit{in vivo} nature of the technique makes time-resolved probing possible. This allows for the measurement of temporal correlation functions -- an observable that is usually not accessible in globally destructive measurements such as optical fluorescence measurements. (ii) The electron beam technique is not restricted to two-dimensional arrays of atoms. It is applicable to any species also when fluorescence imaging is not possible. Multi-atom multi-orbital physics can be addressed as well. (iii) The spatial resolution can in principle be further improved down to a few nanometer. The precise shape of on-site wave functions and small-scale spatial correlations can then be investigated. 

Scanning electron microscopy can also be applied to fermionic gases, where the study of temporal correlation functions, high precision density measurements and non-equilibrium dynamics help to characterize the different quantum phases. However, the magnetic field required for the application of Feshbach resonances can amount to several hundred Gauss and the compatibility with the presented electron microscopy technique is not obvious. The emerging technology of controlling the s-wave scattering length with optical transitions~\cite{Enomoto_2008,Yan_2013,Bauer_2009} provides an alternative strategy for tuning the interactions of a Fermi gas. The recent advent of the dipolar systems erbium \cite{Frisch_2014} and dysprosium \cite{Baumann_2014} exhibit low-field Feshbach resonances that are readily compatible with scanning electron microscopy. 

A very exciting application of scanning electron microsopy is the study of open many-body quantum systems. This includes dissipative many-body models \cite{Diehl_2008, Kordas_2012, Cai_2013, Poletti_2013} and non-Hermitian Hamiltonians \cite{Bender_1998}. Interestingly, non-Hermitian quantum walks have been predicted to show a topological transition \cite{Rudner_2009}. In a more general context, any macroscopic quantum system is influenced by the environment and the interplay between coherence, decoherence and dissipation determines its properties. For instance, recent observations indicate that even biological systems might exploit quantum coherence to enhance their functionality~\cite{Labmert_2013,Huelga_2013}. While decoherence is ubiquitous in nature, its control is challenging. Scanning electron microscopy of ultracold quantum gases allows for a unique way to engineer dissipative processes. This can be used to study generic questions of open quantum systems under idealized conditions, such as dark states, dissipative attractors and quantum phases in open systems.

{\bf Acknowledgements.}
We acknowledge financial support by the DFG within the SFB/TRR 49. We thank all previous and present group members for planning, developing and further advancing this technique.

\section*{References}

\bibliographystyle{unsrt}
\bibliography{references-BS.bib}

\end{document}